\documentclass[twocolumn,twoside,letterpaper,11pt]{article}
\usepackage{graphicx,xrrc,natbib}
\usepackage{times}

\newcommand{\lsimeq}
  {\hbox{\raise0.5ex\hbox{$<\lower1.06ex\hbox{$\kern-1.07em{\sim}$}$}}}
\newcommand{\mnras}{MNRAS}
\newcommand{\apj}{ApJ}
\newcommand{\apjl}{ApJL}
\newcommand{\apjs}{ApJS}

\begin{document}


\title{CONTINUING A CHANDRA SURVEY OF QUASAR RADIO JETS}


%
%
%
%


\author{    J. M. Gelbord, H. L. Marshall                } 
\institute{ Massachusetts Institute of Technology        } 
\address{ 77 Massachusetts Ave., Cambridge, MA 02139, U.S.A. } 
\email{     jonathan@space.mit.edu, hermanm@space.mit.edu} 

\author{D. A. Schwartz, D. M. Worrall, M. Birkinshaw}
\email{das@head-cfa.harvard.edu, D.Worrall@bristol.ac.uk, Mark.Birkinshaw@bristol.ac.uk}
\author{J. E. J. Lovell, D. L. Jauncey, E. S. Perlman}
\email{Jim.Lovell@csiro.au, David.Jauncey@csiro.au, perlman@jca.umbc.edu}
\author{D. W. Murphy, R. A. Preston}
\email{dwm@sgra.jpl.nasa.gov, Robert.A.Preston@jpl.nasa.gov}


\maketitle

\abstract{ We are conducting an X-ray survey of flat spectrum radio
 quasars (FSRQs) with extended radio structures.
 We summarize our results from the first stage of our survey,
 then we present
 findings from its continuation.

 We have discovered jet X-ray emission from 12 of our first 20
 \textit{Chandra} targets,
 establishing that strong 0.5--7.0~keV
 emission is a common feature of FSRQ jets.  The X-ray morphology
 is varied, but in general closely matches the radio structure until
 the first sharp radio bend.
 In the sources with optical data as well as X-ray detections we
 rule out simple synchrotron models for X-ray emission,
 suggesting these systems may instead be dominated by inverse
 Compton (IC) scattering.  Fitting models of IC scattering of
 cosmic microwave background photons suggests that these jets are aligned
 within a few degrees of our line of sight, with bulk Lorentz factors
 of a few to ten and magnetic fields a bit stronger than $10^{-5}$~G.

 In the weeks prior to this meeting, we have discovered two new
 X-ray jets at $z > 1$.  
 One (PKS~B1055+201) has a dramatic, $20''$-long jet.
 The other (PKS~B1421-490) appears unremarkable at radio frequencies,
 but at higher frequencies the jet is uniquely powerful: 
 its optically-dominated, with jet/core flux ratios of 3.7 at 1~keV
 and 380 at 480~nm.}

\section{Introduction}

A quarter century after the discovery of extragalactic X-ray jets
\citep{schreier79}, many fundamental questions remain unanswered.  
Even the process responsible for the high-energy emission remains a
subject of debate: in some instances the radio synchrotron continuum
appears to extend to the X-ray band \citep[e.g., knot A1 of
  3C~273:][]{marshall01}, while in others inverse Compton (IC)
scattering best describes the observations
\citep[e.g., PKS 0637-752:][]{schwartz00}.
In order to assess the distribution of high energy emission mechanisms
amongst quasar jets, and to use this information to determine the range
of physical conditions, we are conducting a
multiwavelength survey of a large sample of flat spectrum quasars
(FSRQs; defined as quasars with core radio spectral index
  values $\alpha < 0.5$, where F$_\nu \propto \nu^{-\alpha}$)
selected by their extended ($> 2''$) flux at 5~GHz \citep{marshall05}.
Brief \textit{Chandra} exposures are used to detect X-ray bright
jets and to identify candidates for follow-up observations; 
radio and optical observations allow multi-waveband morphological
comparisons and fill in the jet spectral energy distribution (SED),
which in turn constrain the emission models. 

\section{The initial survey}
The preliminary survey consists of twenty 5~ks ACIS observations made
during \textit{Chandra} cycle~3.
These X-ray data are supplemented with new
radio maps made with the \textit{ATCA} and \textit{VLA}, and
\textit{Magellan} optical images,
all with subarcsecond resolution.
The twenty targets are drawn from our full sample of 56 FSRQs,
including ten targets from a flux-limited subsample and ten from a
morphologically-selected extension to this sample.

\begin{figure}
  \begin{center}
    \includegraphics[width=0.49\columnwidth]{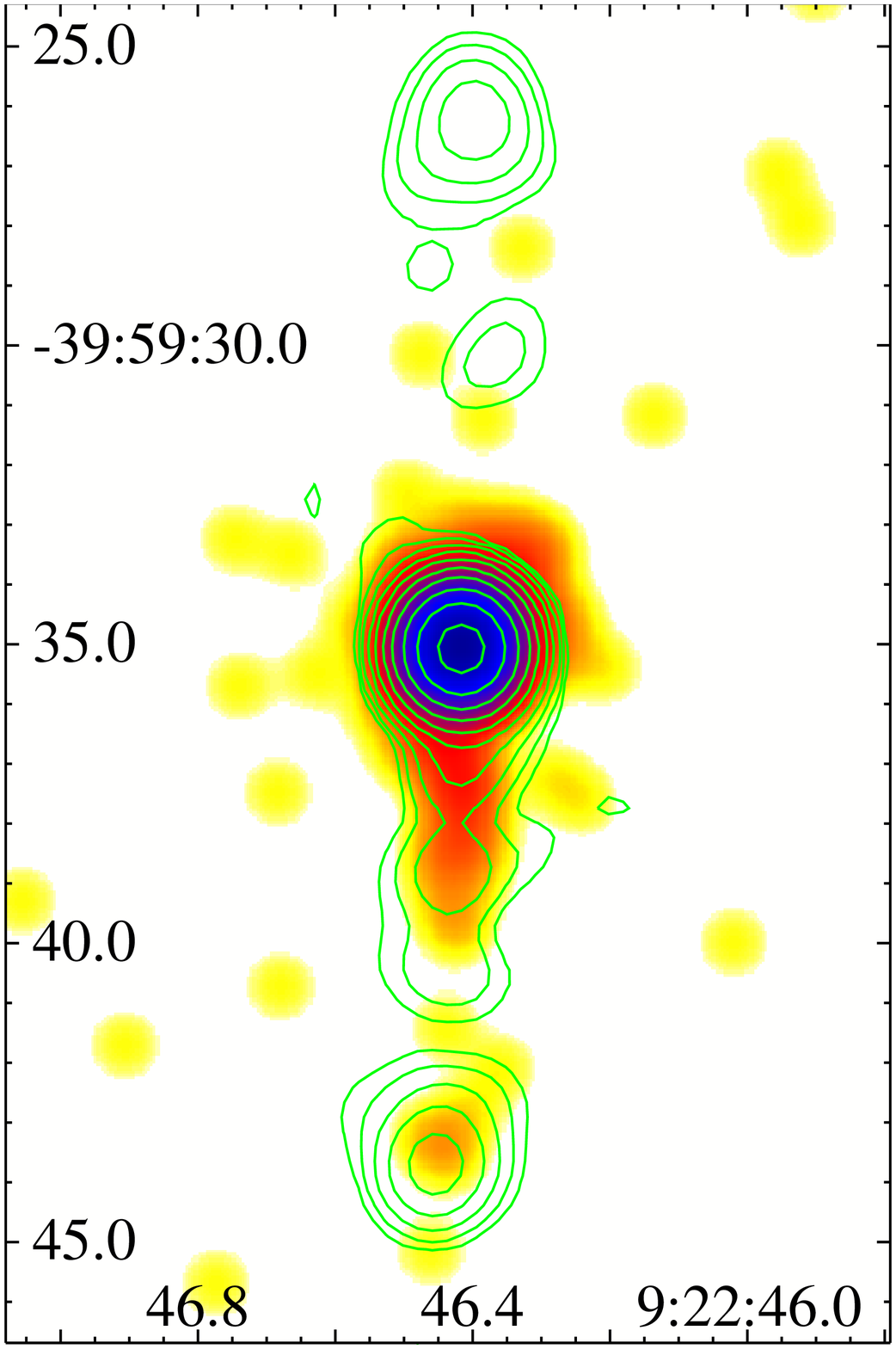}
    \includegraphics[width=0.49\columnwidth]{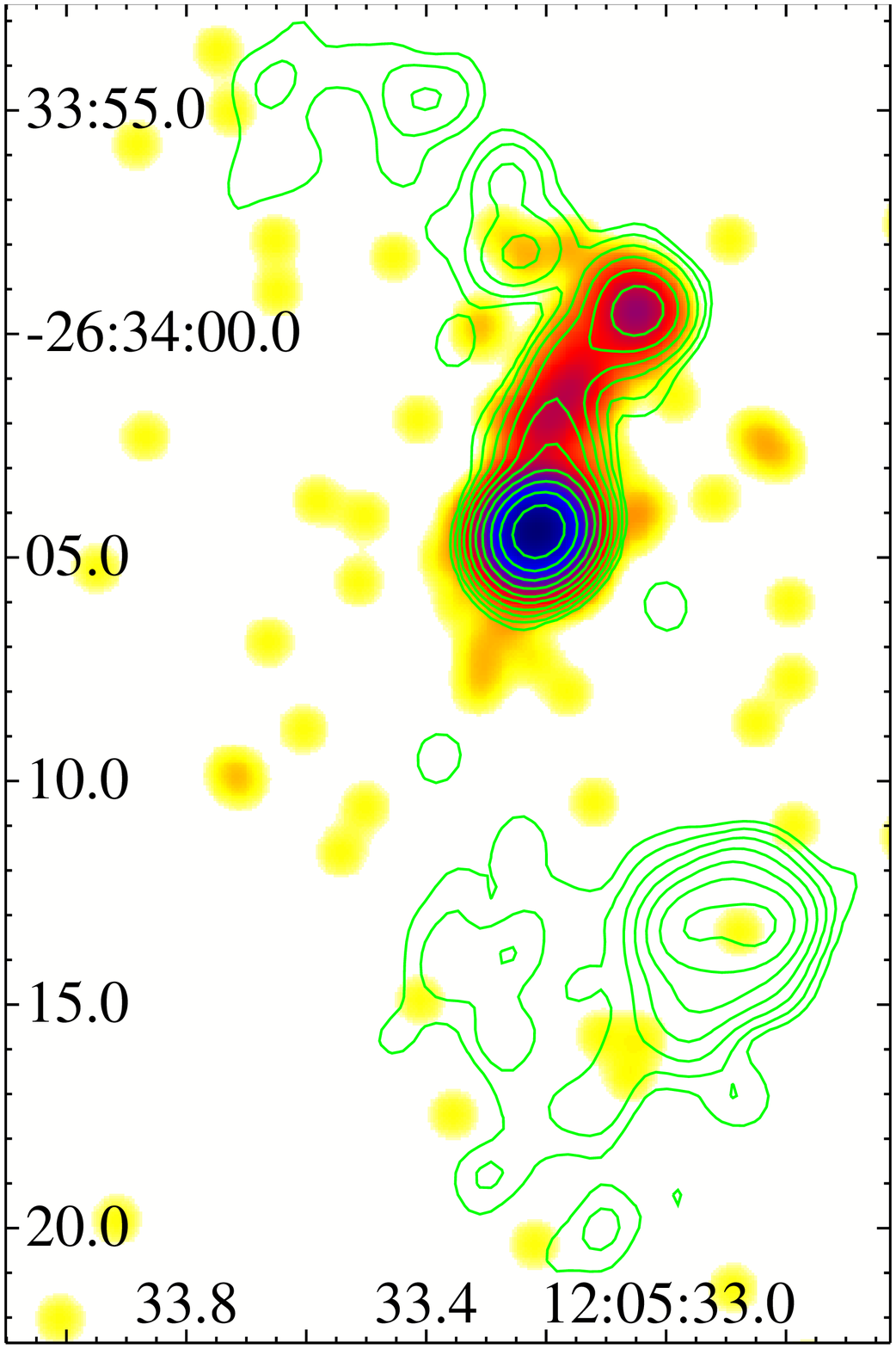}
    \includegraphics[width=\columnwidth]{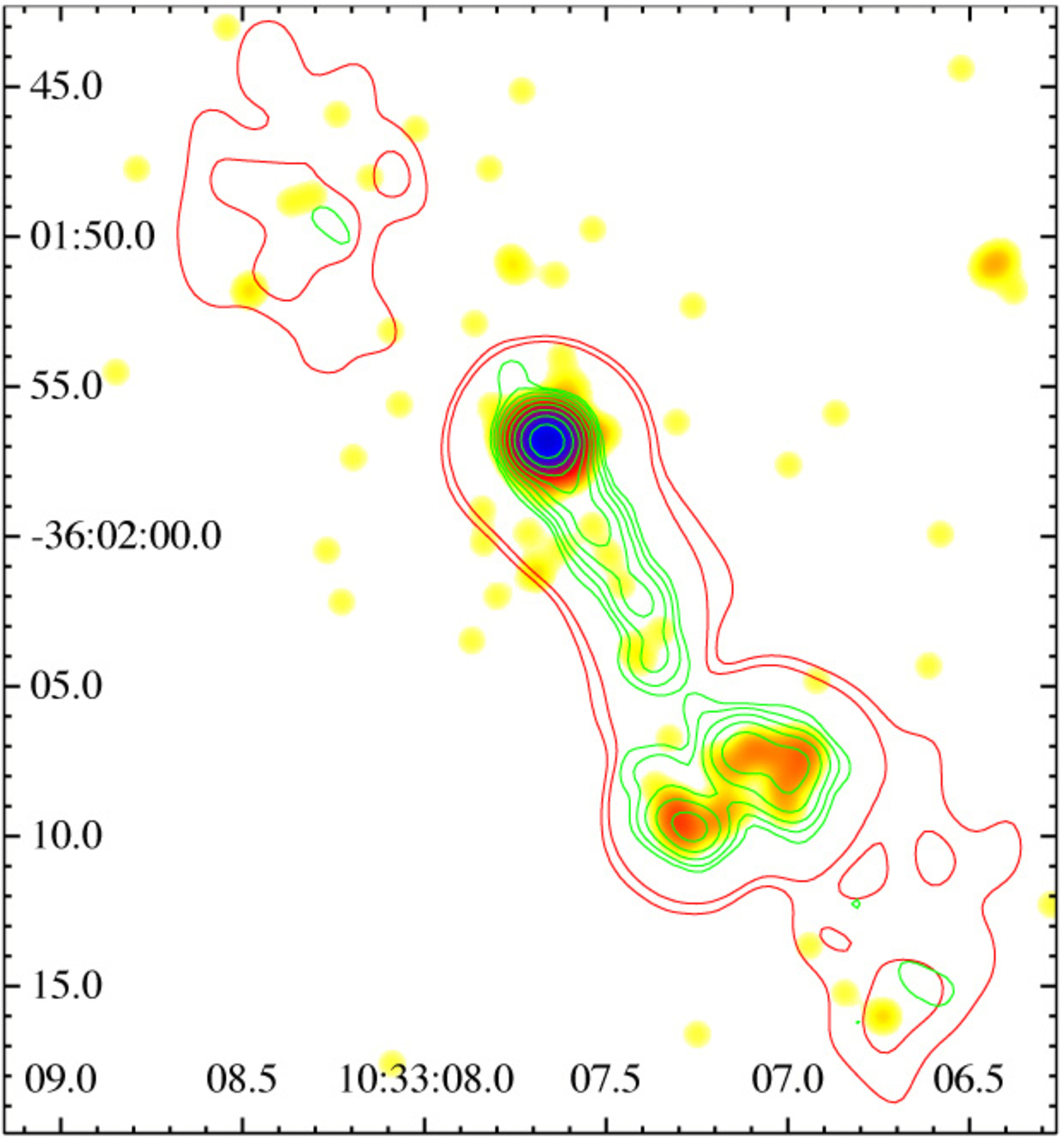}
    \caption{\small \textit{Chandra} 0.5-7.0~keV X-ray images of PKS
      0920-397, PKS 1202-262 and PKS 1030-357 (clockwise from top
      left) with \textit{ATCA} radio contours overlaid.  Images and
      contours have been convolved to matching $1.2''$ FWHM
      resolution.
      All contours start at $5 \times$ the
      background RMS level of the respective radio maps;
      green contours represent 8.6~GHz data and the two outermost
      (red) contours around 1030-357 illustrate the
      larger-scale structure using the lowest levels from the
      4.8~GHz map.
      The upper two images are representative in that the X-rays
      follow the radio jet where it is straight or bends modestly, but
      end where the radio jet bends sharply; the X-ray/radio flux
      ratio declines along the jet of 0920-397 and stays relatively
      constant along the inner $6''$ of 1202-262.
      PKS 1030-357 is unusual in that the X-rays are barely detected
      in the inner jet, intensify at the first knot and remain strong
      through 
      what appears to be two sharp bends in the radio jet.}
    \label{fig:xray}
  \end{center}
\end{figure}

We detect jets in 12 of the \textit{Chandra} sources, indicating
that strong X-ray emission is a common feature of quasar jets.
We consider this discovery rate to be a lower limit
for the incidence of X-ray jets
because the detection rate amongst the flux-limited subsample is
higher (8 out of 10), suggesting that X-ray jets may be present but
below our detection threshold in some of our fainter targets.
All of the detected jets are one-sided, but there is considerable
variety in their details
(Fig.\ \ref{fig:xray}).
X-ray hotspots are found to coincide with radio knots, which is
suggestive of a direct connection between the low- and high-energy
emission mechanisms.
This is consistent with both of the favored models for jet
emission: the X-rays either represent 
the synchrotron emission of the highest-energy electrons,
or IC scattering by the low-energy end of the synchrotron-emitting
electron population.
When the radio jet bends sharply the X-ray jet usually ends or weakens
dramatically, consistent with either a deceleration/depletion of the
highest energy electrons in the synchrotron picture or a change to a
less-favored beaming angle in the IC model \citep{marshall05}.

\begin{figure}
  \begin{center}
    \includegraphics[width=\columnwidth]{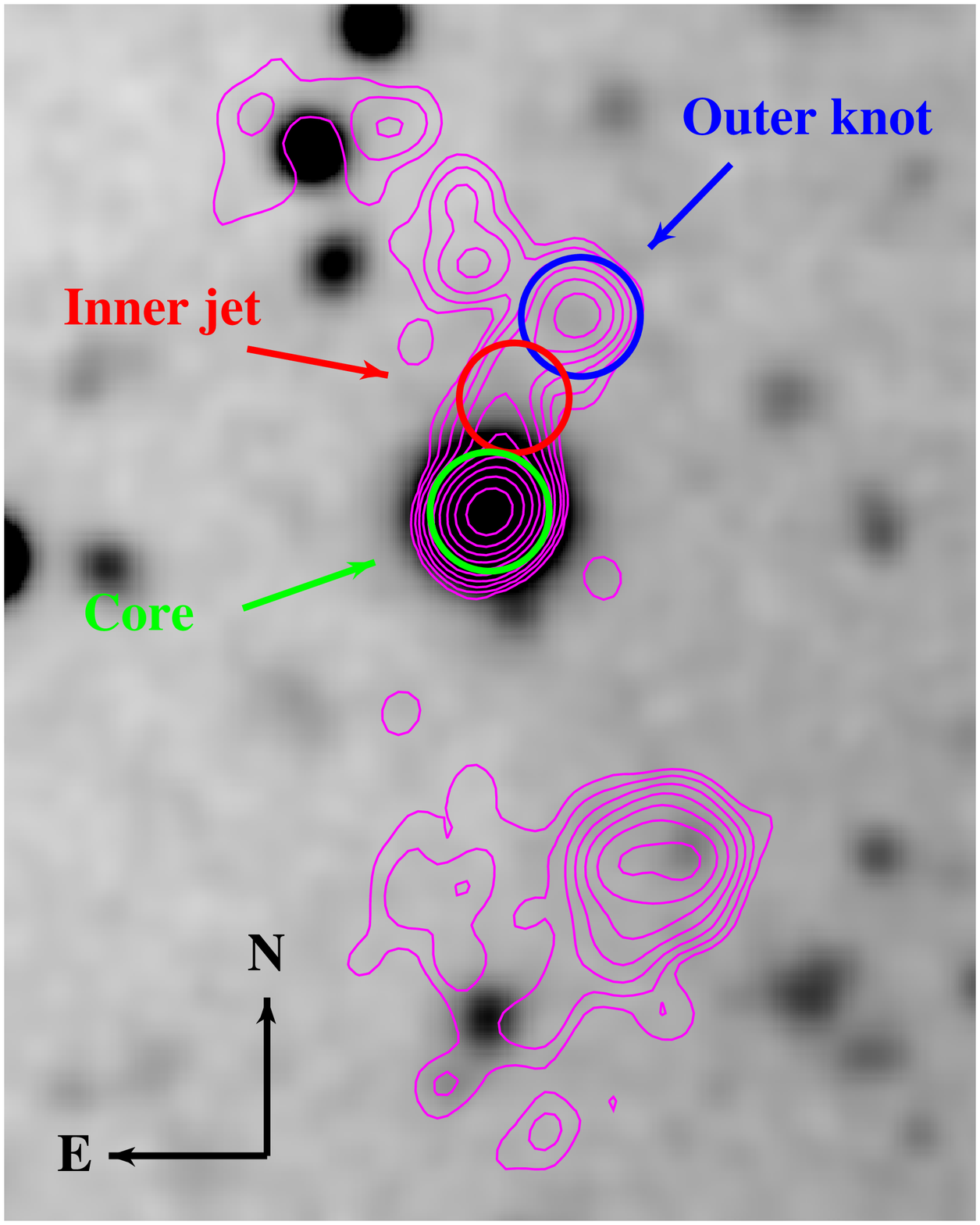}
    \includegraphics[height=\columnwidth,angle=270]{7.14_gelbordFig2balt.eps}
    \caption{\small \textit{Top panel:}
      \textit{Magellan} optical (SDSS $g'$) image of
      PKS~1202-262 with \textit{ATCA} 8.6~GHz radio contours overlaid.  
      \textit{Bottom panel:}
      Radio--optical--X-ray SED plot for the three regions labeled on
      the image.  Simple power law interpolations for the inner jet
      and outer knot are sketched in as dashed and dotted lines,
      respectively.} 
    \label{fig:1202}
  \end{center}
\end{figure}

As of January 2004, only six of the X-ray bright systems had yet been
observed with \textit{Magellan}, and none of these jets were detected.
The optical flux upper limits rule out simple synchrotron
models, which predict a flat or concave-down spectrum unless a second
population of electrons is invoked
(\citealp{harris02}; but see \citealt{dermer02} for an alternative
synchrotron model with a single electron population extending into the
Klein-Nishina regime).  
In the case of PKS~1202-262, the optical fluxes are at least 2--3
magnitudes below the interpolation  between the radio and X-ray fluxes
(Fig.~\ref{fig:1202}), strongly suggesting the IC model.
If we apply jet models of IC scattering off the cosmic microwave
background \citep*{tavecchio00,celotti01}, we obtain magnetic fields of
order $B \sim 10^{-5}$~G, bulk Lorentz factors up to $\Gamma \sim 10$,
kinetic powers of $10^{46}$--$10^{47}$~erg/s, and alignments close to
our line of sight implying deprojected lengths of hundreds of kpc
\citep{schwartz03}.

\section{The latest results}

\subsection{January 2004 \textit{Chandra} observations}

\begin{figure}
  \begin{center}
    \includegraphics[width=\columnwidth]{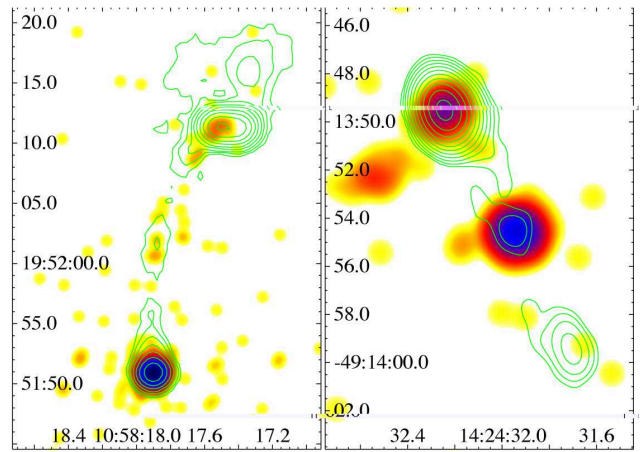}
    \caption{\small \textit{Chandra} 0.5-7.0~keV images of
      PKS~B1055+201 (\textit{left}) and PKS~B1421-490
      (\textit{right});
      the former is shown with 1.4~GHz radio contours, the latter has 8.6~GHz
      contours.}
    \label{fig:newX}
  \end{center}
\end{figure}

Two new sources observed with \textit{Chandra} in the weeks prior to
this meeting have revealed remarkable X-ray structures (Fig.\
\ref{fig:newX}).
The X-rays from PKS~1055+201 (observed 2004 Jan.\ 19) trace a long,
arcing radio jet, terminating at the near side of an extended radio
feature $21.3''$ from the core.
This jet brightens as it approaches the radio feature, possibly
indicating the diffusion of shock accelerated electrons back down the
jet.
PKS~1421-490 (observed 2004 Jan.\ 16) is the first system we've
found to be dominated at high frequencies by its jet.
The unresolved X-ray feature $5.8''$ SW of the radio core provides
79\% of the 0.5--7.0~keV flux.

\subsection{The incredible jet of PKS~1421-490}

\begin{figure}
  \begin{center}
    \includegraphics[width=\columnwidth]{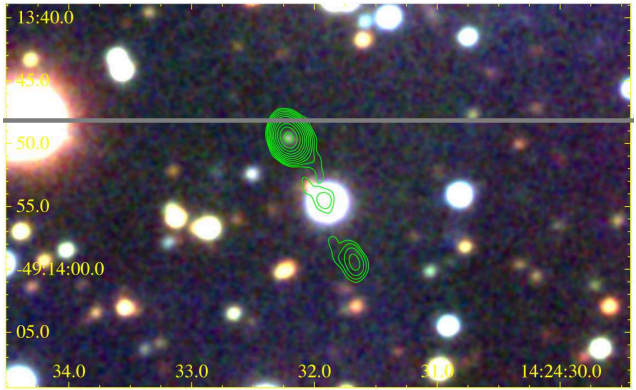}
    \caption{\small \textit{Magellan}
      $g'$-$r'$-$i'$ true-color image of the PKS~B1421-490 field, with
      8.6~GHz radio contours.}
    \label{fig:1421opt}
  \end{center}
\end{figure}

PKS~1421-490 is the only member of our sample without a
previously-identified optical counterpart.
We observed the field with \textit{Magellan} in the SDSS $g'$, $r'$
and $i'$ filters, identifying a 24th magnitude source within $0.3''$
of the radio core (Fig.\ \ref{fig:1421opt}).
The dereddened colors of this source are consistent with quasars at $1
\, \lsimeq \, z \, \lsimeq \, 2$, suggesting that 1421-490 is
one of the
more distant members of our sample \citep{gelbord05b}.

\begin{figure}
  \begin{center}
    \includegraphics[height=\columnwidth,angle=270]{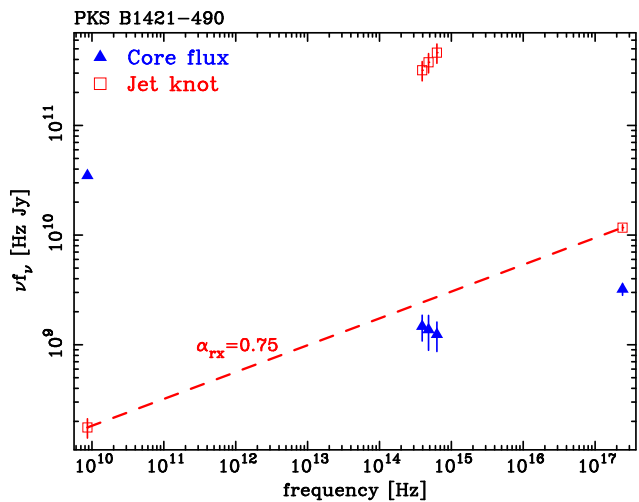}
    \caption{Radio--optical--X-ray SED of the core and
      knot of 1421-490.}
    \label{fig:1421sed}
  \end{center}
\end{figure}

Coinciding with the strong X-ray peak is an unresolved source that is
$\sim$300 times brighter than the optical core.  In no other
known quasar system does the jet so thoroughly overwhelm the core in
the optical band.
This 17th magnitude knot ranks this as the second
brightest extragalactic optical jet component, only slightly fainter
than knot HST-1 of M~87 despite its much greater distance.
The optically-dominated SED of the bright knot (Fig.\ \ref{fig:1421sed}) 
is difficult to
interpret; it may be best explained by IC from a decelerating
relativistic jet aligned close to our line of sight and boosting
downstream photons \citep{georganopoulos03a}.
Another possibility that cannot yet be ruled out is that this ``knot''
is actually an unrelated source.
However, the SED and optical colors combined with a featureless (albeit
low S/N) optical spectrum observed in April 2004 eliminates most
contaminants except for an exotic, optically-dominated BL~Lac object
\citep{gelbord05a}.

\section{On the horizon...}

This survey is very much a work in progress.
A detailed report on the first 20 \textit{Chandra} targets 
is coming out in January \citep{marshall05}.
Of the remaining 36 sources in our sample, half have now been
observed by us or by other investigators.  
Only by conducting large surveys will we discover unusual systems such
as PKS~1421-490.
We now have approved programs for follow-up \textit{Chandra} and
\textit{HST} observations of selected sources.
These data, together with new higher frequency radio observations, will 
provide the necessary data to examine the evolution of properties
along the lengths of the jets through more detailed spatially-resolved SEDs.
Our ground-based optical program is also continuing, with both imaging of
systems not scheduled with \textit{HST} and follow-up spectroscopy for
1421-490 and other sources with ambiguous identifications or unknown
redshifts;
the first collection of our \textit{Magellan} results is due out next
year \citep{gelbord05b}.

\section*{Acknowledgments}

This work has been supported in part under SAO contracts GO4-5124X and
SV1-61010.
The \textit{VLA} is a facility of the National Radio Astronomy
Observatory, operated by Associated Universities, Inc., under
cooperative agreement with the National Science Foundation.
The \textit{ATCA} is part of the Australia Telescope which is
funded by the Commonwealth of Australia for operation as a National
Facility managed by CSIRO.

\end{document}